%% file: main.tex
\documentclass[
superscriptaddress,
reprint, amsmath,amssymb,
pra,
floatfix,
array,
]{revtex4-1}
\usepackage[dvipdfmx]{graphicx}
\usepackage[dvipdfmx]{color}
\usepackage{enumerate}
\usepackage{graphicx}
\usepackage{ascmac}
\input{Qcircuit}

\begin{document}

\title{Interoperability in encoded quantum repeater networks}
\author{Shota Nagayama}
\email{kurosagi@sfc.wide.ad.jp}
\affiliation{Graduate School of Media and Governance, Keio University, 5322 Endo, Fujisawa-shi, Kanagawa 252-0882, Japan}
\author{Byung-Soo Choi}
\affiliation{Research Center for Quantum Information Technology, Electronics and Telecommunications Research Institute, Daejeon, South Korea}
\author{Simon Devitt}
\affiliation{Center for Emergent Matter Science, RIKEN, Wako, Saitama 351-0198, Japan}
\author{Shigeya Suzuki}
\affiliation{Keio Research Institute at SFC, Keio University, 5322 Endo, Fujisawa-shi, Kanagawa 252-0882, Japan}
\author{Rodney Van Meter} 
\affiliation{Faculty of Environment and Information Studies, Keio University, 5322 Endo, Fujisawa-shi, Kanagawa 252-0882, Japan}
\date{\today}
\begin{abstract}
 The future of quantum repeater networking will require interoperability between various error correcting codes.
 A few specific code conversions and even a generalized method are known, however, 
no detailed analysis of these techniques in the context of quantum networking has been performed.
 In this paper, we analyze a generalized procedure to create Bell pairs encoded heterogeneously between two 
 separate codes used often in error corrected quantum repeater network designs.
 We begin with a physical Bell pair, then encode each qubit in a different error correcting code, using entanglement purification to increase the fidelity.
We investigate three separate protocols for preparing the purified encoded Bell pair. 
 We calculate the error probability of those schemes between the Steane [[7,1,3]] code,
 a distance three surface code and single physical qubits by Monte Carlo simulation under a standard Pauli error model,
 and estimate the resource efficiency of the procedures.
 A local gate error rate of $10^{-3}$ allows us to create high-fidelity
logical Bell pairs between any of our chosen codes.
 We find that a postselected model, where any detected parity flips in code stabilizers result in a restart of the protocol, 
 performs the best.
\end{abstract}
\maketitle

\input{introduction}
\input{heterogeneouslyencodedbellpair}

\input{purification}

\input{simulation}

\input{conclusion}
\input{acknowledgement}
\bibliography{shota-reviews.bib}
\newpage
\appendix
\input{data}

\end{document}

%% file: introduction.tex
\section{Introduction}
Much like the Internet of today, it is probable that a future Quantum Internet will be a collection of radically different quantum networks
utilizing some form of quantum inter-networking.
These networks, called {\it Autonomous Systems} in the classical Internet vernacular, are deployed and administered independently,
and realize end-to-end communication by relaying their communication in a technology-independent, distributed fashion for scalability.
In the quantum regime, different error mitigation techniques may be employed within neighboring quantum networks
and a type of code conversion or code teleportation between heterogeneous error correcting codes must be provided for interoperability.

The quantum repeater is a core infrastructure component of a quantum network,
tasked with constructing distributed quantum states or
relaying quantum information as it routes from the source to the destination
~\cite{briegel98:_quant_repeater,dur:PhysRevA.59.169,6246754,kimble08:_quant_internet}.
The quantum repeater creates new capabilities: end-to-end quantum communication,
avoiding limitations on distance and the requirement for trust in quantum key distribution networks~\cite{doi:10.1117/12.606489,1367-2630-11-7-075001,sasaki:11},
wide-area cryptographic functions~\cite{Ben-Or:2005:FQB:1060590.1060662},
distributed computation~\cite{RevModPhys.82.665,buhrman03:_dist_qc,buhrman1998quantum,PhysRevA.89.022317,chien15:_ft-blind,broadbent2010measurement,crepeau:_secur_multi_party_qc}
and possibly use as physical reference frames~\cite{jozsa2000qcs,komar14:_clock_qnet,chuang2000qclk,RevModPhys.79.555}.

Several different classes of quantum repeaters have been proposed~\cite{munro2011designing,vanmeter2014book,takeoka2014fundamental} and these class distinctions often relate to 
how classical information is exchanged when either preparing a connection over multiple repeaters, or sending a piece of quantum information from source to desitination.
The first class utilizes purification and swapping of physical Bell pairs~\cite{Nature:10.1038/35106500,PhysRevLett.96.240501,RevModPhys.83.33,Jiang30102007}.
First, neighboring repeaters establish raw (low fidelity) Bell pairs which are recursively used to purify a single pair to a desired fidelity.  
Adjacent stations then use entanglement swapping protocols to double the total range of the entanglement. In purify/swap protocols, classical information is exchanged 
continuously across the entire network path to herald failures of both purification protocols and entanglement swapping.  This exchange of information limits the speed of such a network 
significantly, especially over long distances.
The second class utilizes quantum error correction throughout the
end-to-end
communication~\cite{munro2010quantum,PhysRevA.79.032325,PhysRevLett.104.180503,1367-2630-15-2-023012} and limits the exchange of classical information to either two-way 
communications between adjacent repeaters or to ballistic communication, where the classical information flow is unidirectional from source to receiver.
These approaches depend on either high probability of success for
transmitting photons over a link with high fidelity, or build on top
of heralded creation of nearest neighbor Bell pairs and
purification, if necessary.  
If the probability of successful connection between adjacent repeaters is high enough we can use quantum
error correcting codes and relax constraints on the
technology, especially memory decoherence times and the need for large
numbers of qubits in individual repeaters, by sending logically
encoded states hop by hop in a quasi-asynchronous
fashion~\cite{munro2010quantum,munro2012quantum} or using
speculative or measurement-based
operations~\cite{1367-2630-15-2-023012,PhysRevA.85.062326,munro2012quantum}.

Independent networks may employ any of the above schemes, and within
some schemes may choose different error correcting codes or code
distances.  Initially deployed to support different applications and
meet technological, logistical, geographic and economic constraints,
they may use different physical implementations and will have
different optimal choices for operational methods.

Over time, however, it will likely become desirable to interconnect
these networks into a single, larger, internetwork.  In this paper, we
address the problem of creating end-to-end entanglement despite
differences at the logical level.

These inter-network and differing operating environments can be bridged through the use of heterogeneously encoded logical Bell pairs,
using separate error correction methods on each half of the pair.
The idea of code conversion goes back a decade for use in large-scale systems.
To achieve interoperability between two QEC codes, we can classify approaches into two groups, direct code conversion and code teleportation.
Direct code conversion transforms an encoded state $\vert \psi \rangle_L$ into an encoded state $\vert \psi \rangle_{L'}$
where $L$ and $L'$ indicate two distinct codes.
Since this change operates on valuable data, the key point is to find an appropriate fault-tolerant sequence that will convert the 
stabilizers from one code to the other
~\cite{PhysRevLett.113.080501,Hill:2013:FQE:2481614.2481619,PhysRevA.77.062335}.
In code teleportation, conversion is achieved by teleporting information using a heterogeneously encoded Bell pair as 
a resource state. Therefore, the key point is the method for preparing such a state.

\begin{figure}[t]
 \begin{center}
  \includegraphics[width=8.5cm]{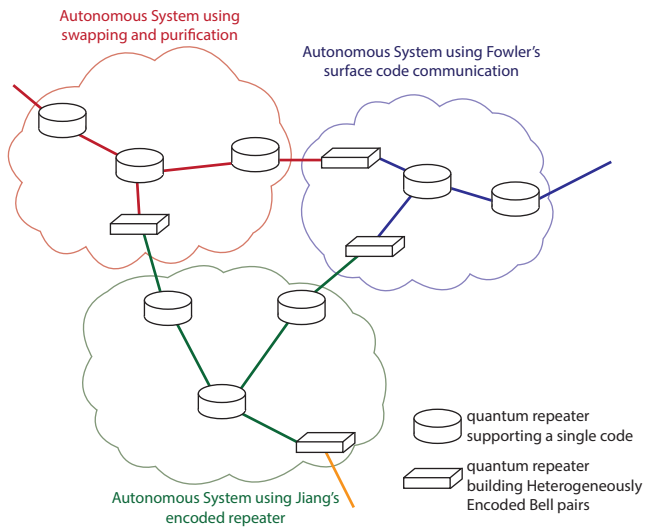}
  \caption{A use case for quantum repeaters building heterogeneously encoded Bell pairs.
  Each cloud represents a quantum {\it autonomous system} which is based on an error correcting code
  or entanglement swapping and purification.
  Colored links are connections using those codes.
  Boxes are quantum repeaters building heterogeneously encoded Bell pairs.
  Cylinders are quantum repeaters, each of which supports only a single code.
  All links from a homogeneous repeater (cylinder) are the same type (color)
  since only quantum repeaters building heterogeneously encoded Bell pairs (boxes) can interoperate between different codes.
  }
  \label{fig:use_case}
 \end{center}
\end{figure}

Figure \ref{fig:use_case} shows an example use case for heterogeneously encoded Bell pairs, used in quantum {\it autonomous systems}
\footnote{In the classical Internet, significant differences may occur
even between subnets of a single AS, but for simplicity in this paper
we will restrict ourselves to the assumption that a quantum AS is
internally homogeneous.}.
Quantum {\it autonomous systems} of different codes interoperate via quantum repeaters building heterogeneously encoded Bell pairs.

\begin{figure*}[t]
 \begin{center}
  \includegraphics[width=17cm]{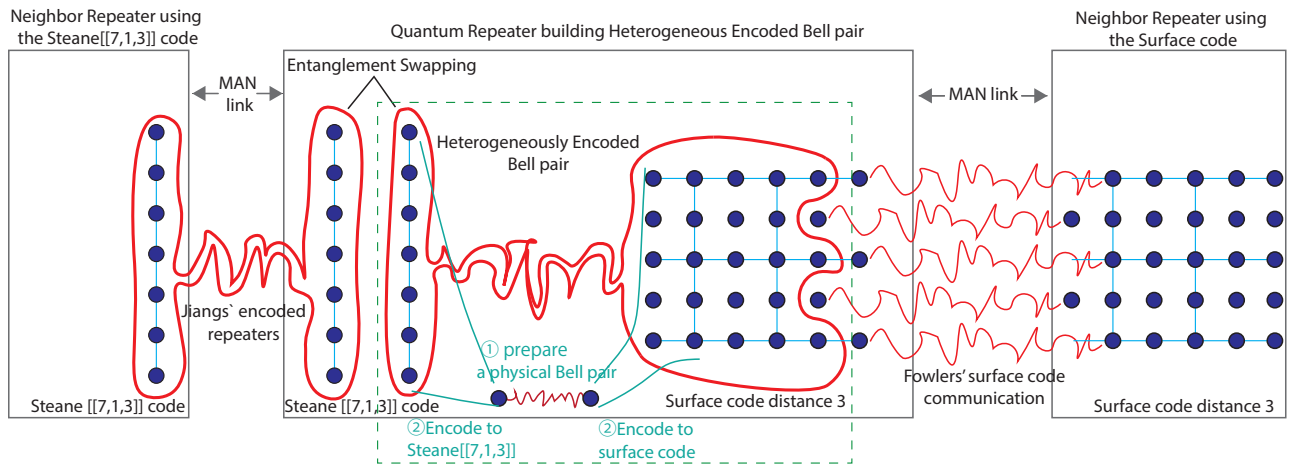}
  \caption{Heterogeneously encoded Bell pairs can be used to bridge quantum networks using different error correcting mechanisms.
  MAN stands for metropolitan area network.
  A blue dot denotes a physical qubit. A set of blue lines indicates qubits which comprise an encoded qubit.
  Each thin red line describes an entanglement between physical qubits.
  Each thick red loop outlines entanglement between encoded qubits.
  The half of the Bell pair encoded in the surface code can be sent to the neighboring quantum repeater by Fowler \emph{et al.}'s method
  ~\cite{PhysRevLett.104.180503}.
  The other half of the Bell pair, encoded in the Steane [[7,1,3]] code, undergoes entanglement swapping
  with a Steane [[7,1,3]]-Steane [[7,1,3]]] encoded Bell pair
 established via Jiang \emph{t al.}'s method~\cite{PhysRevA.79.032325}.
  Therefore this central quantum repeater can create entanglement between the two quantum repeaters in different types of networks.
  In the green dashed rectangle is the procedure for encoding a Bell pair heterogeneously.
  A qubit of a Bell pair is encoded onto Steane [[7,1,3]] on the left side of the figure, adding 6 qubits.
  The other qubit of the Bell pair is encoded onto the surface code of distance 3, adding 24 qubits
  on the right side of the figure. Multiple copies are prepared, entangled and purified.  
  Eventually, a heterogeneously encoded logical Bell pair
  is achieved with high enough fidelity to enable coupling of the two networks.}
  \label{fig:use_in_internet}
 \end{center}
\end{figure*}

In this paper, we give  the first detailed analysis of the generalized approach to create heterogeneously encoded Bell pairs for interoperability of quantum error correcting networks.
We evaluate this approach between the Steane [[7,1,3]] code, a distance three surface code,
and unencoded (raw) physical qubits.
Figure \ref{fig:use_in_internet} depicts a quantum repeater building and using heterogeneously encoded Bell pairs to be used in a quantum repeater.

We have studied three possible schemes to increase the fidelity of the heterogeneously encoded Bell pairs:
{\it purification before encoding}, {\it purification after encoding} and {\it purification after encoding with strict post-selection}.
{\it Purification before encoding} does entanglement purification at the level of physical Bell pairs.
{\it Purification after encoding} does entanglement purification at the level of encoded Bell pairs.
{\it Purification after encoding with strict post-selection} also does entanglement purification at the level of encoded Bell pairs.
The difference from the previous scheme is that encoded Bell pairs in which any eigenvalue (error syndrome) of -1 is measured
in the purification stage are discarded and the protocols restarted.
We determine the error probability and the resource efficiency of these schemes by Monte Carlo simulation
with the Pauli error model of circuit level noise~\cite{landahl:arXiv:1108.5738}.

%% file: heterogeneouslyencodedbellpair.tex
\section{Heterogeneously encoded Bell pairs}
There are two methods for building heterogeneously encoded Bell pairs for code teleportation.
The first is to inject each qubit of a physical Bell pair to a different code
~\cite{copsey02:_quant-mem-hier}.
The second is to prepare a common cat state for two codes to check the ZZ parity of two logical qubits
~\cite{976922,Thaker:2006:QMH:1150019.1136518,doi:10.2200/S00331ED1V01Y201101CAC013}.
It has been shown that code teleportation utilizing a cat state is better than direct 
code conversion because the necessary stabilizer checking for the latter approach is too expensive~\cite{6675561}.
Direct code conversion and code teleportation utilizing a cat state are specific for a chosen code pair as the specific 
sequence of fault-tolerant operations has to match the two codes chosen.
In contrast, code teleportation by injecting a physical Bell pair can be used for any two codes, and provided encoding circuits 
are available for the two codes in question, the protocol can be generalized to arbitrary codes.

Putting things together, heterogeneous Bell pairs of long distance can be created
by entanglement swapping (physical or logical) or a method appropriate to each network, allowing
an arbitrary quantum state encoded in some code 
to be moved onto another code by teleportation
~\cite{PhysRevA.79.032325,PhysRevLett.104.180503}.
In a single computer, code conversion has been proposed for memory hierarchies and for cost-effective fault tolerant quantum computation
~\cite{copsey02:_quant-mem-hier,Thaker:2006:QMH:1150019.1136518,PhysRevLett.113.080501,PhysRevLett.112.010505,PhysRevLett.111.090505,choi:dual_code_model}.

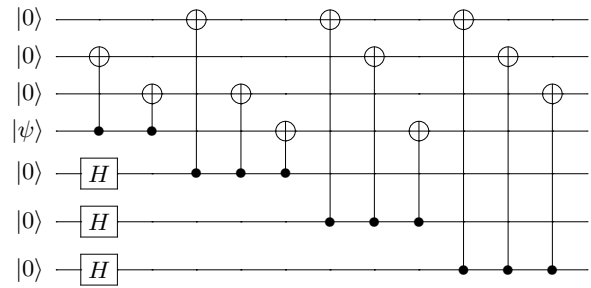
\begin{figure}[t]
 \begin{center}
\[
     \Qcircuit @C=1em @R=.7em {
& \lstick{\ket{0}}   & \qw         & \qw       & \targ    &\qw         & \qw    &\targ      &\qw        & \qw     & \targ    &\qw           & \qw       & \qw \\
& \lstick{\ket{0}}   & \targ       & \qw       & \qw    &\qw           & \qw    &\qw        & \targ     &\qw        &\qw         & \targ     & \qw       & \qw \\
& \lstick{\ket{0}}   & \qw         & \targ     & \qw       & \targ     &\qw      &\qw       &\qw        & \qw       & \qw    &\qw           & \targ     & \qw \\
& \lstick{\ket{\psi}}& \ctrl{-2}   & \ctrl{-1} & \qw    &\qw           & \targ   & \qw      &\qw        &\targ         & \qw     & \qw    &\qw           & \qw \\
& \lstick{\ket{0}}   & \gate{H}    & \qw       & \ctrl{-4} & \ctrl{-2} &\ctrl{-1}& \qw      &\qw        &\qw     & \qw       & \qw    &\qw           & \qw \\
& \lstick{\ket{0}}   & \gate{H}    & \qw       & \qw    &\qw           & \qw    &\ctrl{-5}  & \ctrl{-4} &\ctrl{-2}     & \qw & \qw    &\qw           & \qw \\
& \lstick{\ket{0}}   & \gate{H}    & \qw       & \qw    &\qw           & \qw    &\qw        &\qw        & \qw    &\ctrl{-6}     & \ctrl{-5} & \ctrl{-4} & \qw \\
     }
\]
  \caption{Circuit to encode an arbitrary state to the Steane [[7,1,3]] code~\cite{Sidney:10.1007:s11128-012-0414-7}.
  $\vert \psi \rangle$ is the state to be encoded.
  This circuit is not fault-tolerant. The KQ of this circuit is 42 because some gates can be performed simultaneously.}
  \label{fig:CSS7qubitEncode}
 \end{center}
\end{figure}

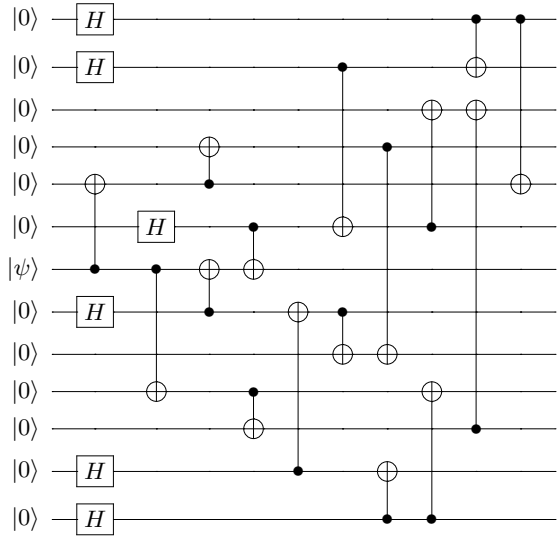
\begin{figure}[t]
 \begin{center}
    \[
    \Qcircuit @C=1em @R=.7em {
& \lstick{\ket{0}}  & \gate{H} & \qw & \qw & \qw & \qw & \qw & \qw & \qw & \ctrl{1} & \ctrl{4} & \qw \\
& \lstick{\ket{0}}  & \gate{H} & \qw & \qw & \qw & \qw & \ctrl{4}& \qw &\qw& \targ& \qw & \qw \\
& \lstick{\ket{0}}  & \qw & \qw & \qw & \qw & \qw & \qw & \qw & \targ& \targ& \qw & \qw \\
& \lstick{\ket{0}}  & \qw & \qw & \targ&\qw& \qw & \qw & \ctrl{5} &\qw& \qw & \qw & \qw \\
& \lstick{\ket{0}}  & \targ&\qw& \ctrl{-1}&\qw& \qw & \qw & \qw & \qw & \qw & \targ& \qw \\
& \lstick{\ket{0}}  & \qw & \gate{H} & \qw & \ctrl{1}& \qw & \targ& \qw & \ctrl{-3}& \qw & \qw & \qw \\
& \lstick{\ket{\psi}}  & \ctrl{-2}& \ctrl{3}& \targ& \targ& \qw & \qw & \qw & \qw & \qw & \qw & \qw \\
& \lstick{\ket{0}}  & \gate{H} & \qw & \ctrl{-1}& \qw & \targ& \ctrl{1}& \qw & \qw & \qw & \qw & \qw \\
& \lstick{\ket{0}}  & \qw & \qw & \qw & \qw & \qw & \targ & \targ&\qw& \qw & \qw & \qw \\
& \lstick{\ket{0}}  & \qw & \targ&\qw& \ctrl{1}& \qw & \qw & \qw & \targ&\qw& \qw & \qw \\
& \lstick{\ket{0}}  & \qw & \qw & \qw & \targ& \qw & \qw & \qw & \qw & \ctrl{-8}& \qw & \qw \\
& \lstick{\ket{0}}  & \gate{H} & \qw & \qw & \qw & \ctrl{-4}&\qw& \targ &\qw& \qw & \qw & \qw \\
    & \lstick{\ket{0}}  & \gate{H} & \qw & \qw & \qw & \qw & \qw & \ctrl{-1} & \ctrl{-3}&\qw&\qw& \qw \\
    }
    \]
  \caption{Circuit to encode an arbitrary state $\vert \psi \rangle$ to a distance three surface code
  ~\cite{:/content/aip/journal/jmp/43/9/10.1063/1.1499754}.
  This circuit is not fault-tolerant.
  The KQ of this circuit is 250 if some gates are performed simultaneously.
  }
  \label{fig:SurfaceCodeDist3Encode}
 \end{center}
\end{figure}

The green dashed rectangle in Figure \ref{fig:use_in_internet} shows the basic procedure for creating
a heterogeneously encoded logical Bell pair. Each dot denotes a
physical qubit and thin blue lines connecting those dots demark the set of physical qubits
comprising a logical qubit.
Each qubit of a Bell pair is processed separately and encoded onto its respective code
through non-fault-tolerant methods to create arbitrary encoded states.
Figure \ref{fig:CSS7qubitEncode} shows the circuit to encode an arbitrary quantum state
in the Steane[[7,1,3]] code~\cite{steane:10.1098/rspa.1996.0136,steane:10.1038/20127}.
Figure \ref{fig:SurfaceCodeDist3Encode} shows the circuit to encode an arbitrary quantum state
in the surface code~\cite{:/content/aip/journal/jmp/43/9/10.1063/1.1499754}. 
The KQ of a circuit is the number of qubits times the circuit depth, giving an estimate of the number of opportunities for errors to occur~\cite{steane:10.1098/rspa.1996.0136}.
Note that those circuits are not required to be fault-tolerant because the state being purified is generic,
rather than inreplaceable data.
If the fidelity of the encoded Bell pair is not good enough
(e.g. as determined operationally using quantum state tomography),
entanglement purification is performed~\cite{bravyi2005uqc,PhysRevA.73.062309}.

%% file: purification.tex
 \section{Three Methods to Prepare a Heterogeneously Encoded High Fidelity Bell Pair}
Entanglement purification is performed to establish high fidelity entanglement
~\cite{duer03:_pur-qc,dur2007epa}.
Entanglement purification can be viewed as a distributed procedure for testing a proposition about a distributed state~\cite{vanmeter2014book}.

\begin{figure}[t]
 \begin{center}
\[
\Qcircuit @C=1em @R=.7em {
&            && \ru& \ctrl{1} & \gate{H} &\qw &  &\\
& \ket{\phi}&& \ru& \targ    & \meter   &     & \rd \ru   &\ket{\phi''} \\
& \ket{\phi'}&& \rd& \ctrl{1} & \gate{H} &\qw &  &\\
&            && \rd& \targ    & \meter   &    &     & \\
}
\]
 \caption{Circuit for entanglement purification~\cite{dur:PhysRevA.59.169}.
 The two measured values are compared.
  If they disagree, the output qubits are discarded.
  If they agree, the output qubits are treated as a new Bell pair.
  At this point, the X error rate of the output Bell pair is suppressed from the input Bell pairs.
  The Hadamard gates exchange the X and Z axes, so that the following round of purification suppresses the Z error rate.
  As the result, entanglement purification consumes two Bell pairs and
  generates a Bell pair of higher fidelity stochastically.
 }
 \label{sc:circ_purification}
 \end{center}
\end{figure}
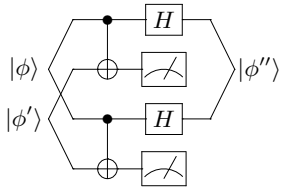
Figure \ref{sc:circ_purification} shows the circuit for the basic form of entanglement purification where $\vert\phi\rangle$ is a noisy Bell pair.
The input is two low fidelity Bell pairs and on success the output is a Bell pair of higher fidelity.
 One round of purification suppresses one type of error, X or Z.
If the initial Bell pairs are Werner states, or approximately Werner states, then to suppress both types, two rounds of purification are required.
The first round makes the resulting state into a binary state with only one significant error term but not a significantly improved fidelity.
The second round then strongly suppresses errors if the gate error rate is small.
Thus, the overall fidelity tends to improve in a stair step fashion.
After two rounds of purification, the distilled fidelity will be, in the absence of local gate error,
\begin{equation}
 F'' \sim \frac{F^2}{F^2 + (1-F)^2}
\end{equation}
where the original state is the Werner state
\begin{equation}
 \rho = F|\Phi^+\rangle\langle\Phi^+| + \frac{1-F}{3}(|\Phi^-\rangle\langle\Phi^-| + |\Psi^+\rangle\langle\Psi^+|+ |\Psi^-\rangle\langle\Psi^-|)
\end{equation}
and $F$ is the fidelity $F=\langle\phi|\rho|\phi\rangle$ if $|\phi\rangle$ is the desired state.
The probability of success of a round of purification is
\begin{equation}
 p = F^2 + 2F\frac{1-F}{3} + 5\left(\frac{1-F}{3}\right)^2.
\end{equation}
Table \ref{tab:purificationresult} in the appendix provides the numerical data for this to compare with our protocols.
Our simulation assumptions are detailed in section \ref{sec:sim}.

\subsection{{\it Purification before encoding}}
\begin{figure}[t]
  \begin{center}
   \includegraphics[width=9cm]{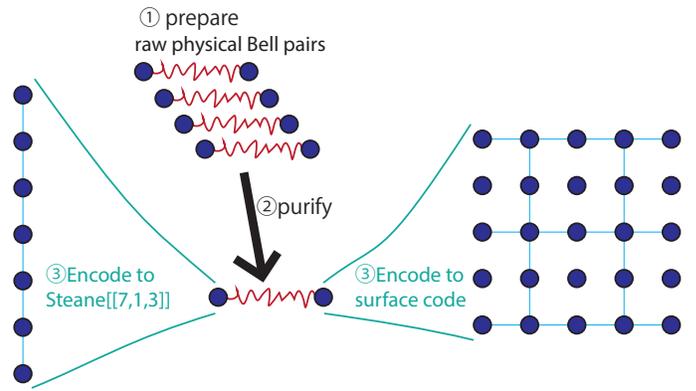}
   \caption{Overview of the scheme which purifies physical Bell pairs to generate an encoded Bell pair of high fidelity.
   First, entanglement purification is conducted between physical Bell pairs an arbitrary number of times.
   Second, each qubit of the purified physical Bell pair is encoded to heterogeneous error correcting code.}
   \label{fig:overview_purify_before_encode}
  \end{center}
\end{figure}
Figure \ref{fig:overview_purify_before_encode} shows the overview of the scheme to make heterogeneously encoded Bell pairs
that are purified before encoding.
To create an encoded Bell pair of high fidelity,
entanglement purification is repeated the desired number of times.
Next, each qubit of the purified Bell pair is encoded to its respective error correcting code.
To estimate the rate of {\em logical} error after encoding, we perform a perfect 
syndrome extraction of the system to remove any residual correctable errors.
After the whole procedure finishes, we check whether logical errors exist.
 Table \ref{tab:purify_before_encode} in the appendix presents the details of the simulated error probability and resource efficiency of {\it purification before encoding}.

\subsection{{\it Purification after encoding}}
\begin{figure}[t]
  \begin{center}
   \includegraphics[width=9cm]{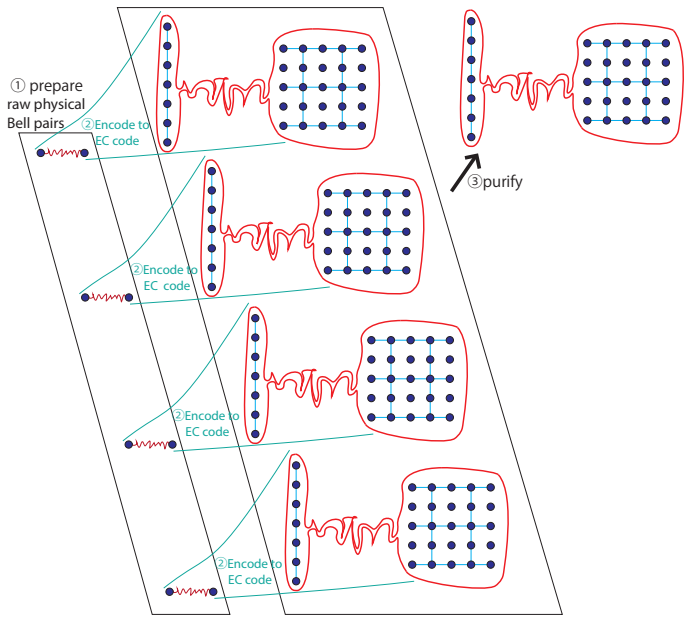}
   \caption{Overview of the scheme which purifies encoded Bell pairs to achieve an encoded Bell pair of high fidelity.
   In this method, first, raw physical Bell pairs are encoded into our heterogeneous error correcting code,
   Secondly, those heterogeneously encoded Bell pairs are purified
   directly at the logical level.}
   \label{fig:overview_purify_after_encode}
  \end{center}
\end{figure}
Figure \ref{fig:overview_purify_after_encode} shows the overview of the scheme to make heterogeneously encoded Bell pairs
that are purified after encoding.
In this scheme, to create an encoded Bell pair of high fidelity,
heterogeneously encoded Bell pairs are generated first by encoding each qubit of a raw physical Bell pair
to our chosen heterogeneous error correcting codes.
Next, those encoded Bell pairs are purified at the logical level the desired number of times, via transversal CNOTs and logical measurements.
 Table \ref{tab:purify_after_encode} presents the details of the simulated error probability and resource efficiency of {\it purification after encoding}.

\subsection{{\it Purification after encoding with strict post-selection}}
\begin{figure}[t]
  \begin{center}
   \includegraphics[width=9cm]{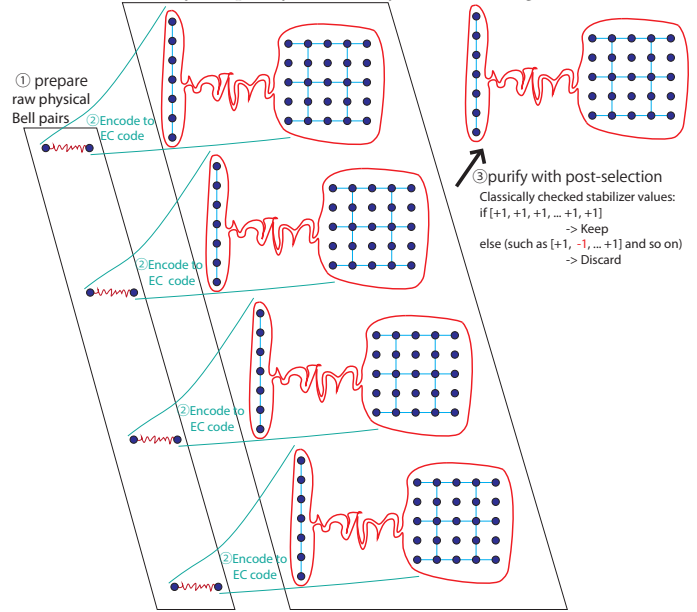}
   \caption{Overview of the scheme which purifies encoded Bell pairs to achieve an encoded Bell pair of high fidelity with
   strict post-selection. First, raw physical Bell pairs are encoded to heterogeneous error correcting code, same as {\it purification after encoding}.
   Secondly, at measurement in purification, eigenvalues of each stabilizer are checked classically.
   If any eigenvalue of -1 is measured, the output Bell pair is discarded (in a similar manner to if the overlying purification 
   protocol failed).
   }
   \label{fig:overview_purify_after_encode_strict}
  \end{center}
\end{figure}
Figure \ref{fig:overview_purify_after_encode_strict} shows the overview of the scheme to make encoded Bell pairs,
purified after encoding with strict post-selection protocols to detect errors.
This scheme uses a procedure similar to {\it purification after encoding}.
In this scheme, to create an encoded Bell pair of high fidelity,
heterogeneously encoded Bell pairs are generated first by encoding each qubit of a raw physical Bell pair
to our chosen heterogeneous error correcting codes. We then run purification protocols at the logical level, similarly to the previous 
protocol. However, when we perform a logical measurement as part of this protocol, we also calculate (classically) the 
eigenvalues of all code stabilizers. If any of these eigenvalues are found to be negative, we treat the operation as a failure
(in a similar manner to odd parity logical measurements for the purification)
and the output Bell pair of the purification is discarded. This simultaneously performs purification and 
error correction using the properties of the codes.
 Table \ref{tab:purify_after_encode_strict} presents the details of the numerically calculated error probability and
 resource efficiency of {\it purification after encoding with strict post-selection}.

%% file: simulation.tex
\section{Error Simulation and Resource Analysis}
\label{sec:sim}
We calculate the error probability and estimate resource requirements by Monte Carlo simulation.
The physical Bell pairs' fidelity is assumed to be 0.85; the state is assumed to be, following N\"olleke \emph{et al.}~\cite{PhysRevLett.110.140403},
 \begin{widetext}
\begin{equation}
\rho = 0.85 |\Phi^+\rangle\langle\Phi^+| + 0.04 |\Phi^-\rangle\langle\Phi^-| + 0.055|\Psi^+\rangle\langle\Psi^+| + 0.055|\Psi^-\rangle\langle\Psi^-|.
  \label{equ:input_bell_pair}
\end{equation}
 \end{widetext}

Our error model is the Pauli model of circuit level noise~\cite{landahl:arXiv:1108.5738}.
This model consists of memory error, 1-qubit gate error, 2-qubit gate error, and measurement error
each of which occurs with the error probability $p$.
Memory, 1-qubit gates and measurement are all vulnerable to X, Y and Z errors and we assume a balanced model, where probabilities are $\frac{p}{3}$ respectively.
Similarly, 2-qubit gates are vulnerable to all fifteen possibilities, 
each with a probability of $\frac{p}{15}$.
Errors propagate during all circuits after the initial distribution of Bell pairs.

\begin{figure*}[t]
  \begin{center}
   \includegraphics[width=18cm]{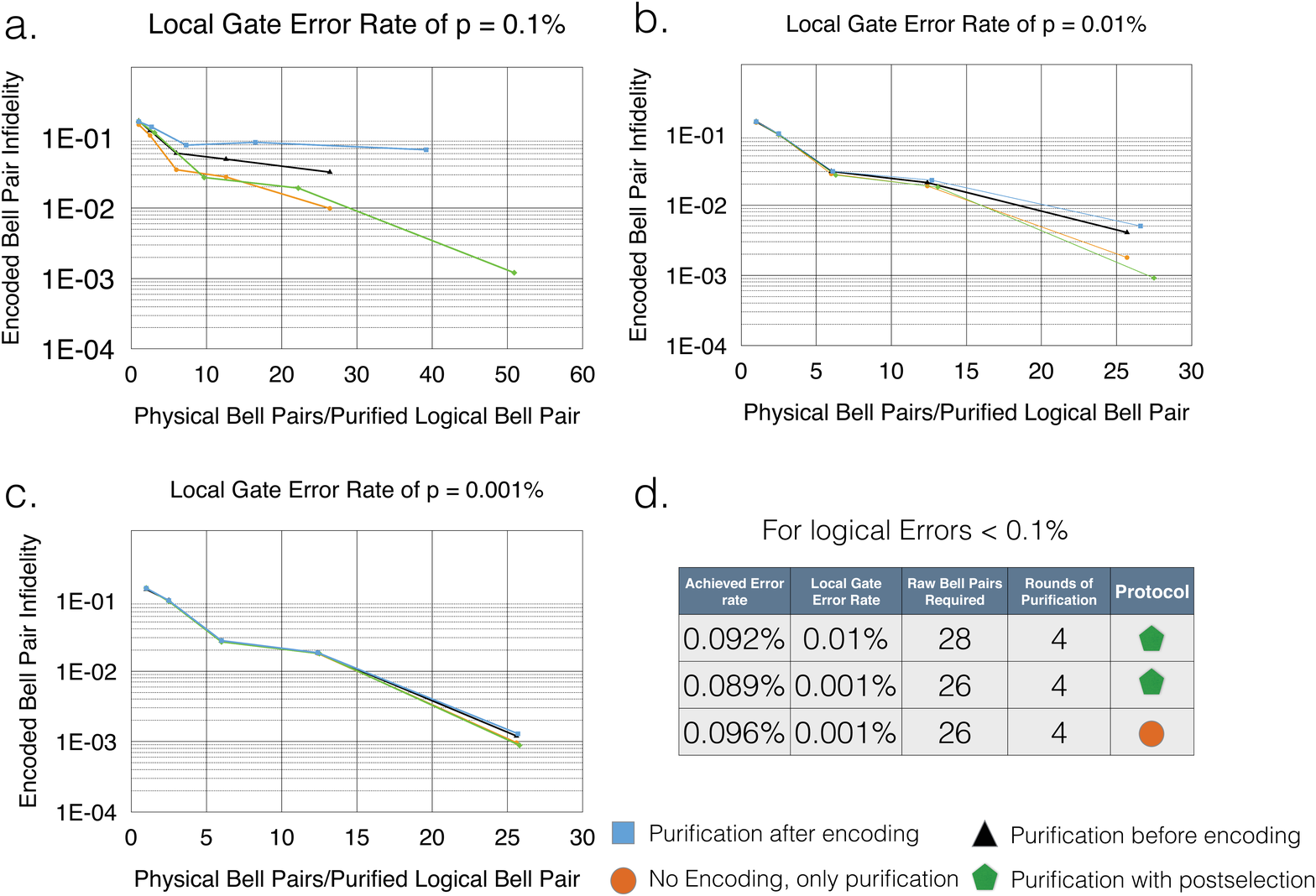}
\caption{Results of simulation of creation of a Steane
   [[7,1,3]]-surface code distance 3 heterogeneous Bell pair, showing
   residual logical error rate versus physical Bell pairs consumed.
   The three schemes plus the baseline case of purification of
   physical Bell pairs are each represented by a line.  Each point
   along a line corresponds to the number of rounds of purification.
   The leftmost point represents no purification, the second point is
   one round of purification, and the rightmost point represents four
   rounds of purification a.-c. Improving values of
   local gate error rate.  d.  The three cases with residual error
   rate of $10^{-3}$ or less.}
   \label{fig:graph_resource_vs_pl}
  \end{center}
\end{figure*}

 Figure \ref{fig:graph_resource_vs_pl} plots the numbers of consumed raw physical Bell pairs versus logical error rate
 in the output state.

The numbers of raw Bell pairs consumed declines as the local gate error rate is lowered.
This is because the influence of the local gate error rate shrinks relative to the infidelity of generated raw Bell pairs.
If the system is free from local gate error, the numbers of raw Bell pairs consumed by the three schemes must converge.
At $p=10^{-5}$, the required number of raw Bell pairs of the schemes are essentially identical and
they require about 26 raw Bell pairs to achieve four rounds of purification.
Higher efficiency would require improving the initial fidelity of $F=0.85$.

At any error rate and with any number of rounds of purification from 0 to 4,
{\it purification before encoding} and {\it purification after encoding}
result in fidelity worse than simple purification of physical Bell pairs.
This suggests that errors accumulated during encoding are difficult to correct.
On the other hand, {\it purification after encoding with strict post-selection} 
gives better results than simple purification, at the expense of consuming more raw physical Bell pairs.
This difference is noticeable at $p=10^{-3}$; 51 raw physical Bell pairs are used
to create an encoded Bell pair purified four rounds.
The local gate error rate is so high that an eigenvalue of -1 is often found
at the measurement in purification and the output Bell pair is discarded.
For {\it purification after encoding with post-selection}, the residual error rate after $n$ rounds of purification
is similar at any $p$, but resource demands change. It converts local errors into ``loss'', or discarded states.
Therefore {\it purification after encoding with post-selection} is dominated by the original raw Bell pair infidelity.
At $p=10^{-3}$, {\it purification after encoding} also requires more raw physical Bell pairs than the other schemes, because the error rate after purification is so high that
the success probability of purification is poor.

Though more rounds of purification are supposed to result in smaller logical error rate, 
three rounds of purification of {\it purification after encoding} at $p=10^{-3}$ give an error rate {\it larger}
than that of two rounds. The local gate error rate is too high and
purification introduces more errors than it suppresses on odd-numbered purification rounds.

{\it Purification after encoding with strict post-selection}
gives similar results for the two local gate error rates $p=10^{-4}$ and $p=10^{-5}$.
The difference is a small number of consumed raw physical Bell pairs.
From this fact we conclude that $p=10^{-3}$ is a good enough local gate error rate
to allow us to create heterogeneously encoded Bell pairs from raw physical Bell pairs of F=0.85.

%% file: conclusion.tex
\section{Discussion}
We have proposed and analysed a generalized method for creating heterogeneously encoded Bell pairs that 
can be used for interoperability
between encoded networks.  This is the first step in examining the full design of 
interconnection routers for quantum repeater systems utilizing different error mitigation techniques.  
Our results have shown that {\it purification after encoding with strict post-selection} is a better preparation method than our other two candidates.
Strict post-selection of two rounds of purification results in better fidelity
than error correction of four rounds of purification at all error rates,
and better physical Bell pair efficiency.
Since the threshold of the error rate of the Steane [[7,1,3]] code is around $10^{-4}$,
our simulations of {\it purification before encoding} and
{\it purification after encoding} of $\sim 10^{-4}$ do not show an advantage compared to simple physical purification;
however, strict post-selection does.
{\it Purification after encoding with strict post-selection} has a higher threshold than the normal encoding and purification do.
With initial $F=0.85$, we can reach error rate of $10^{-3}$ (good enough for distributed numeric computation)
using 4 rounds of purification, for physical Bell pairs at $p=10^{-5}$ or
post-selected heterogeneous pairs at $p=10^{-4}$.

As we noted in the introduction, quantum repeater networks will serve
several purposes, potentially requiring different residual error
levels on the end-to-end quantum communication.  Networks using
physical purify-and-swap technologies, for example, will easily
support QKD, but distributed numeric computation will require building
error correction on top of the Bell pairs provided by the network.
Our simulations of heterogeneous Bell pairs where one half is a
physical qubit, rather than logically encoded, are described in the
appendix.  These simulations show that residual error rates can be
suppressed successfully, allowing us to bridge these separate types of
networks and support the deployment of any application suitable for
purify-and-swap networks across a heterogeneous quantum Internet.  The
error rates we have achieved for each heterogeneous technology pair
demonstrate the effectiveness of our heterogeneous scheme for
interoperability.  Moreover, operation appears to be feasible at a
local gate error rate of $10^{-3}$, and at $10^{-4}$ operation is
almost indistinguishable from having perfect local gates.

The analysis presented here is useful not only in the abstract, but
also serves as a first step toward a hardware design for a
multi-protocol quantum router (the boxes in Figure~\ref{fig:use_case}).
Such a router may be built on a quantum multicomputer architecture,
with several small quantum computers coupled internally via a local optical
network~\cite{van-meter10:dist_arch_ijqi,jiang07:PhysRevA.76.062323,kim09:_integ_optic_ion_trap,oi06:_dist-ion-trap-qec}.
This allows hardware architects to build separate,
small devices to connect to each type of network, then to create Bell
pairs between these devices using the method described in this paper.
In addition, this method can be used within large-scale quantum
computers that wish to use different quantum error correcting codes
for different purposes, such as long-term memory or ancilla state
preparation.

This scheme is internal to a single repeater at the border of two networks, and will allow
effective end-to-end communication where errors across links are more important than errors within a repeater node.
It therefore can serve as a building block for a quantum Internet.

%% file: acknowledgement.tex
\section*{acknowledgement}
This work is supported by JSPS KAKENHI Grant Number 25280034 and 25:4103.
We thank Joe Touch for valuable technical conversations.

%% file: data.tex
\section{Detailed Data}
\label{data}
Table \ref{tab:purificationresult} shows our baseline simulation results using physical entanglement only with no encoding.
Table \ref{tab:purify_before_encode} shows the simulated results of {\it purification before encoding} for a Bell pair of a single layer of the Steane [[7,1,3]] code and a distance 3 surface code.
Table \ref{tab:purify_after_encode} shows the simulated results of the scheme {\it purification after encoding} of the same codes.
Table \ref{tab:purify_after_encode_strict} shows the simulated results of the scheme {\it purification after encoding with strict post-selection}.
Since purification at the level of encoded qubits consists of logical gates, {\it purification before encoding} has a much smaller KQ than the other two schemes.
{\it Purification after encoding with strict post-selection} discards more qubits than {\it purification after encoding} does to create a purified encoded Bell pair, so that {\it purification after encoding with strict post-selection} also results in a larger KQ.
Table \ref{tab:purify_after_encode_strict_css_phys} shows the simulated results of the scheme {\it purification after encoding with strict post-selection} between the Steane [[7,1,3]] code and the non-encoded physical half.
Table \ref{tab:purify_after_encode_strict_surface_phys} shows the simulated results of the scheme {\it purification after encoding with strict post-selection} between the distance three surface code and the non-encoded physical half.
   \begin{center}
 \begin{table*}
   \begin{center}
    \caption{Our baseline case, discrete simulation using physical entanglement purification only.
    The merged error rate is the probability that either X error or Z error occurs.
    The physical Bell pair inefficiency is $(\#\ created\ raw\ Bell\ pairs)/(\#\ purified\ Bell\ pairs)$.
    KQ is $\#qubit \times \#steps$. In this simulation, KQ is the number of chances that errors may occur.
    }
    \label{tab:purificationresult}
    (a)The local gate error rate is $10^{-3}$.\\
    \begin{tabular}[t]{|l||l|l|l|r|r|r|r|}
     \hline
     \#purification&X error rate &Z error rate & Merged error rate & Phys. Bell Pair Ineff. & KQ & \#single qubit gate & \#two qubit gate\\
     \hline
0 & 0.112 & 0.0999 & 0.154 & 1.0 & 88 & 86 & 1 \\
1 & 0.0979 & 0.0201 & 0.108 & 2.5 & 98 & 91 & 5 \\
2 & 0.0248 & 0.0145 & 0.0352 & 6.0 & 122 & 103 & 14 \\
3 & 0.0251 & 0.00501 & 0.0278 & 12.6 & 167 & 126 & 32 \\
     4 & 0.0073 & 0.00491 & 0.00993 & 26.4 & 262 & 173 & 70 \\
     \hline
    \end{tabular}
    
    (b)The local gate error rate is $10^{-4}$.\\
    \begin{tabular}[t]{|l||l|l|l|r|r|r|r|}
     \hline
     \#purification&X error rate &Z error rate & Merged error rate & Phys. Bell Pair Ineff. & KQ & \#single qubit gate & \#two qubit gate\\
     \hline
0 & 0.11 & 0.096 & 0.15 & 1.0 & 88 & 86 & 1 \\
1 & 0.0927 & 0.0159 & 0.101 & 2.5 & 98 & 91 & 5 \\
2 & 0.0187 & 0.011 & 0.0278 & 6.0 & 121 & 103 & 14 \\
3 & 0.0182 & 0.000791 & 0.0187 & 12.4 & 166 & 125 & 32 \\
     4 & 0.00126 & 0.000758 & 0.00179 & 25.7 & 258 & 171 & 68 \\
     \hline
    \end{tabular}
    
     (c)The local gate error rate is $10^{-5}$.\\
    \begin{tabular}[t]{|l||l|l|l|r|r|r|r|}
     \hline
     \#purification&X error rate &Z error rate & Merged error rate & Phys. Bell Pair Ineff. & KQ & \#single qubit gate & \#two qubit gate\\
     \hline
0 & 0.109 & 0.0961 & 0.15 & 1.0 & 88 & 86 & 1 \\
1 & 0.0926 & 0.0153 & 0.1 & 2.5 & 98 & 91 & 5 \\
2 & 0.0178 & 0.0102 & 0.0264 & 6.0 & 121 & 103 & 14 \\
3 & 0.0176 & 0.000369 & 0.0179 & 12.4 & 166 & 125 & 32 \\
     4 & 0.000635 & 0.000353 & 0.000963 & 25.6 & 257 & 171 & 68 \\
     \hline
    \end{tabular} 
   \end{center}
 \end{table*}
   \end{center}
 
   \begin{center}
 \begin{table*}
   \begin{center}
    \caption{Simulation results of {\it purification before encoding} for a Bell pair of a single layer of the
    Steane [[7,1,3]] code and a distance 3 surface code.
    Other conditions and definitions are as in Table \ref{tab:purificationresult}.
    }
    \label{tab:purify_before_encode}
    (a)The local gate error rate is $10^{-3}$.\\
    \begin{tabular}[t]{|l||l|l|l|r|r|r|r|}
     \hline
     \#purification&X error rate &Z error rate & Merged error rate & Phys. Bell Pair Ineff. & KQ & \#single qubit gate & \#two qubit gate\\
     \hline
0 & 0.121 & 0.108 & 0.173 & 1.0 & 5624 & 4328 & 648 \\
1 & 0.106 & 0.0354 & 0.127 & 2.5 & 5634 & 4333 & 652 \\
2 & 0.0365 & 0.0314 & 0.0605 & 6.0 & 5658 & 4345 & 661 \\
3 & 0.0346 & 0.0206 & 0.0497 & 12.6 & 5703 & 4368 & 679 \\
4 & 0.0181 & 0.0191 & 0.0325 & 26.4 & 5798 & 4415 & 717 \\    \hline
    \end{tabular}
    
    (b)The local gate error rate is $10^{-4}$.\\
    \begin{tabular}[t]{|l||l|l|l|r|r|r|r|}
     \hline
     \#purification&X error rate &Z error rate & Merged error rate & Phys. Bell Pair Ineff. & KQ & \#single qubit gate & \#two qubit gate\\
     \hline
0 & 0.112 & 0.0962 & 0.154 & 1.0 & 5624 & 4328 & 648 \\
1 & 0.0927 & 0.0171 & 0.102 & 2.5 & 5634 & 4333 & 652 \\
2 & 0.02 & 0.0123 & 0.0302 & 6.0 & 5658 & 4345 & 661 \\
3 & 0.0192 & 0.00224 & 0.0209 & 12.4 & 5702 & 4367 & 679 \\
     4 & 0.00232 & 0.00222 & 0.00407 & 25.7 & 5794 & 4413 & 715 \\
     \hline
    \end{tabular}
    
     (c)The local gate error rate is $10^{-5}$.\\
    \begin{tabular}[t]{|l||l|l|l|r|r|r|r|}
     \hline
     \#purification&X error rate &Z error rate & Merged error rate & Phys. Bell Pair Ineff. & KQ & \#single qubit gate & \#two qubit gate\\
     \hline
0 & 0.109 & 0.0913 & 0.146 & 1.0 & 5624 & 4328 & 648 \\
1 & 0.0927 & 0.0156 & 0.101 & 2.5 & 5634 & 4333 & 652 \\
2 & 0.0179 & 0.0106 & 0.0269 & 6.0 & 5657 & 4345 & 661 \\
3 & 0.0177 & 0.000552 & 0.0182 & 12.4 & 5702 & 4367 & 679 \\
     4 & 0.000745 & 0.000497 & 0.0012 & 25.6 & 5793 & 4413 & 715 \\
     \hline
    \end{tabular} 
   \end{center}
 \end{table*}
  \end{center}
  
   \begin{center}
 \begin{table*}
   \begin{center}
    \caption{Simulation results of the scheme {\it purification after encoding}
    between the Steane [[7,1,3]] code and the distance three surface code.
    Other conditions and definitions are as in Table \ref{tab:purificationresult}.
    }
    \label{tab:purify_after_encode}
    (a)The local gate error rate is $10^{-3}$.\\
    \begin{tabular}[t]{|l||l|l|l|r|r|r|r|}
     \hline
    \#purification&X error rate &Z error rate & Merged error rate & Phys. Bell Pair Ineff. & KQ & \#single qubit gate & \#two qubit gate\\
    \hline
0 & 0.122 & 0.108 & 0.17 & 1.0 & 5624 & 4328 & 648 \\
1 & 0.126 & 0.032 & 0.143 & 2.7 & 7516 & 6034 & 756 \\
2 & 0.0448 & 0.0441 & 0.0787 & 7.3 & 12730 & 10735 & 1052 \\
3 & 0.0744 & 0.0172 & 0.0862 & 16.5 & 23118 & 20099 & 1644 \\
     4 & 0.0341 & 0.0378 & 0.0676 & 39.2 & 48667 & 43129 & 3100 \\
     \hline
    \end{tabular} 

    (b)The local gate error rate is $10^{-4}$.\\
    \begin{tabular}[t]{|l||l|l|l|r|r|r|r|}
     \hline
    \#purification&X error rate &Z error rate & Merged error rate & Phys. Bell Pair Ineff. & KQ & \#single qubit gate & \#two qubit gate\\
    \hline
0 & 0.113 & 0.097 & 0.154 & 1.0 & 5624 & 4328 & 648 \\
1 & 0.0958 & 0.0168 & 0.104 & 2.5 & 7334 & 5868 & 746 \\
2 & 0.0194 & 0.0128 & 0.0302 & 6.1 & 11427 & 9554 & 981 \\
3 & 0.0215 & 0.00147 & 0.0226 & 12.7 & 18984 & 16357 & 1416 \\
4 & 0.00257 & 0.0027 & 0.00503 & 26.6 & 34861 & 30650 & 2329 \\
     \hline
    \end{tabular} 

    (c)The local gate error rate is $10^{-5}$.\\
    \begin{tabular}[t]{|l||l|l|l|r|r|r|r|}
     \hline
    \#purification&X error rate &Z error rate & Merged error rate & Phys. Bell Pair Ineff. & KQ & \#single qubit gate & \#two qubit gate\\
    \hline
0 & 0.111 & 0.0946 & 0.151 & 1.0 & 5624 & 4328 & 648 \\
1 & 0.0938 & 0.0151 & 0.101 & 2.5 & 7311 & 5847 & 745 \\
2 & 0.018 & 0.0108 & 0.0272 & 6.0 & 11294 & 9433 & 973 \\
3 & 0.0179 & 0.000441 & 0.0183 & 12.4 & 18642 & 16047 & 1397 \\
4 & 0.000769 & 0.000547 & 0.00129 & 25.7 & 33870 & 29755 & 2274 \\
     \hline
    \end{tabular} 
   \end{center}
 \end{table*}
  \end{center}

   \begin{table*}
   \begin{center}
    \caption{Simulation results of the scheme {\it purification after encoding with strict post-selection}
    between the Steane [[7,1,3]] code and the distance three surface code.
    Other conditions and definitions are as in Table \ref{tab:purificationresult}.
    }
    \label{tab:purify_after_encode_strict}
    (a)The local gate error rate is $10^{-3}$.\\
    \begin{tabular}[t]{|l||l|l|l|r|r|r|r|}
     \hline
    \#purification&X error rate &Z error rate & Merged error rate & Phys. Bell Pair Ineff. & KQ & \#single qubit gate & \#two qubit gate\\
    \hline
0 & 0.118 & 0.115 & 0.171 & 1.0 & 5624 & 4328 & 648 \\
1 & 0.11 & 0.0159 & 0.118 & 3.1 & 7932 & 6410 & 778 \\
2 & 0.0179 & 0.011 & 0.0273 & 9.7 & 15249 & 13011 & 1189 \\
3 & 0.019 & 0.000398 & 0.0193 & 22.2 & 29090 & 25492 & 1969 \\
4 & 0.000798 & 0.000423 & 0.00121 & 50.9 & 61048 & 54314 & 3769 \\
     \hline
    \end{tabular} 

    (b)The local gate error rate is $10^{-4}$.\\
    \begin{tabular}[t]{|l||l|l|l|r|r|r|r|}
     \hline
    \#purification&X error rate &Z error rate & Merged error rate & Phys. Bell Pair Ineff. & KQ & \#single qubit gate & \#two qubit gate\\
    \hline
0 & 0.112 & 0.0993 & 0.154 & 1.0 & 5624 & 4328 & 648 \\
1 & 0.0935 & 0.0147 & 0.101 & 2.5 & 7366 & 5897 & 748 \\
2 & 0.0179 & 0.0102 & 0.0267 & 6.3 & 11592 & 9702 & 990 \\
3 & 0.0177 & 0.000327 & 0.018 & 13.1 & 19397 & 16730 & 1438 \\
4 & 0.000575 & 0.000343 & 0.000915 & 27.5 & 35733 & 31437 & 2377 \\
     \hline
    \end{tabular} 

    (c)The local gate error rate is $10^{-5}$.\\
    \begin{tabular}[t]{|l||l|l|l|r|r|r|r|}
     \hline
    \#purification&X error rate &Z error rate & Merged error rate & Phys. Bell Pair Ineff. & KQ & \#single qubit gate & \#two qubit gate\\
    \hline
0 & 0.111 & 0.0983 & 0.152 & 1.0 & 5624 & 4328 & 648 \\
1 & 0.0915 & 0.0144 & 0.0983 & 2.5 & 7318 & 5853 & 745 \\
2 & 0.0176 & 0.00998 & 0.0261 & 6.0 & 11304 & 9442 & 974 \\
3 & 0.0175 & 0.000321 & 0.0178 & 12.5 & 18687 & 16088 & 1399 \\
4 & 0.000571 & 0.000317 & 0.000886 & 25.8 & 33957 & 29833 & 2279 \\
     \hline
    \end{tabular} 
\end{center}
 \end{table*}

   \begin{table*}
   \begin{center}
    \caption{Simulation results of the scheme {\it purification after encoding with strict post-selection}
    between the Steane [[7,1,3]] code and non-encoded physical half.
    Other conditions and definitions are as in Table \ref{tab:purificationresult}.
    }
    \label{tab:purify_after_encode_strict_css_phys}
    (a)The local gate error rate is $10^{-3}$.\\
    \begin{tabular}[t]{|l||l|l|l|r|r|r|r|}
     \hline
    \#purification&X error rate &Z error rate & Merged error rate & Phys. Bell Pair Ineff. & KQ & \#single qubit gate & \#two qubit gate\\
    \hline
0 & 0.112 & 0.106 & 0.161 & 1.0 & 4260 & 3660 & 300 \\
1 & 0.103 & 0.0168 & 0.111 & 2.7 & 5370 & 4712 & 330 \\
2 & 0.0213 & 0.0133 & 0.0324 & 6.9 & 8235 & 7426 & 409 \\
3 & 0.0229 & 0.00141 & 0.0238 & 14.9 & 13687 & 12590 & 558 \\
4 & 0.00283 & 0.00149 & 0.00382 & 32.4 & 25534 & 23813 & 883 \\
     \hline
    \end{tabular}
    
    (b)The local gate error rate is $10^{-4}$.\\
    \begin{tabular}[t]{|l||l|l|l|r|r|r|r|}
     \hline
    \#purification&X error rate &Z error rate & Merged error rate & Phys. Bell Pair Ineff. & KQ & \#single qubit gate & \#two qubit gate\\
    \hline
0 & 0.109 & 0.0945 & 0.149 & 1.0 & 4260 & 3660 & 300 \\
1 & 0.0926 & 0.0156 & 0.101 & 2.5 & 5283 & 4629 & 328 \\
2 & 0.0182 & 0.0103 & 0.027 & 6.1 & 7706 & 6923 & 395 \\
3 & 0.018 & 0.000419 & 0.0183 & 12.6 & 12204 & 11181 & 518 \\
4 & 0.000792 & 0.000448 & 0.00119 & 26.2 & 21554 & 20033 & 775 \\
     \hline
    \end{tabular} 

    (c)The local gate error rate is $10^{-5}$.\\
    \begin{tabular}[t]{|l||l|l|l|r|r|r|r|}
     \hline
    \#purification&X error rate &Z error rate & Merged error rate & Phys. Bell Pair Ineff. & KQ & \#single qubit gate & \#two qubit gate\\
    \hline
0 & 0.109 & 0.0969 & 0.15 & 1.0 & 4260 & 3660 & 300 \\
1 & 0.0927 & 0.015 & 0.0999 & 2.5 & 5275 & 4621 & 328 \\
2 & 0.0181 & 0.0105 & 0.0271 & 6.0 & 7656 & 6875 & 393 \\
3 & 0.0175 & 0.000334 & 0.0178 & 12.4 & 12069 & 11053 & 515 \\
     4 & 0.000572 & 0.000325 & 0.000895 & 25.7 & 21202 & 19700 & 766 \\
     \hline
    \end{tabular} 
\end{center}
 \end{table*}

    \begin{table*}
   \begin{center}
    \caption{Simulation results of the scheme {\it purification after encoding with strict post-selection}
    between the distance three surface code and non-encoded physical half.
    Other conditions and definitions are as in Table \ref{tab:purificationresult}.
    }
    \label{tab:purify_after_encode_strict_surface_phys}
    (a)The local gate error rate is $10^{-3}$.\\
    \begin{tabular}[t]{|l||l|l|l|r|r|r|r|}
     \hline
    \#purification&X error rate &Z error rate & Merged error rate & Phys. Bell Pair Ineff. & KQ & \#single qubit gate & \#two qubit gate\\
    \hline
0 & 0.124 & 0.109 & 0.172 & 1.0 & 1296 & 998 & 149 \\
1 & 0.111 & 0.018 & 0.12 & 2.9 & 2303 & 1859 & 237 \\
2 & 0.023 & 0.0137 & 0.0342 & 8.2 & 5180 & 4318 & 489 \\
3 & 0.0233 & 0.00148 & 0.0243 & 18.2 & 10665 & 9006 & 970 \\
4 & 0.0029 & 0.00152 & 0.00392 & 40.8 & 22995 & 19543 & 2053 \\
     \hline
    \end{tabular}
    
    (b)The local gate error rate is $10^{-4}$.\\
    \begin{tabular}[t]{|l||l|l|l|r|r|r|r|}
     \hline
    \#purification&X error rate &Z error rate & Merged error rate & Phys. Bell Pair Ineff. & KQ & \#single qubit gate & \#two qubit gate\\
    \hline
0 & 0.113 & 0.0976 & 0.153 & 1.0 & 1296 & 998 & 149 \\
1 & 0.0946 & 0.0153 & 0.102 & 2.5 & 2131 & 1711 & 223 \\
2 & 0.018 & 0.0102 & 0.0265 & 6.2 & 4147 & 3434 & 402 \\
3 & 0.018 & 0.00046 & 0.0184 & 12.9 & 7865 & 6608 & 734 \\
4 & 0.000798 & 0.000433 & 0.00118 & 26.8 & 15626 & 13236 & 1426 \\
     \hline
    \end{tabular} 

    (c)The local gate error rate is $10^{-5}$.\\
    \begin{tabular}[t]{|l||l|l|l|r|r|r|r|}
     \hline
    \#purification&X error rate &Z error rate & Merged error rate & Phys. Bell Pair Ineff. & KQ & \#single qubit gate & \#two qubit gate\\
    \hline
0 & 0.109 & 0.0928 & 0.151 & 1.0 & 1296 & 998 & 149 \\
1 & 0.093 & 0.0145 & 0.1 & 2.5 & 2120 & 1702 & 222 \\
2 & 0.0179 & 0.01 & 0.0263 & 6.0 & 4059 & 3358 & 395 \\
3 & 0.0176 & 0.00034 & 0.0179 & 12.4 & 7636 & 6413 & 714 \\
4 & 0.000589 & 0.000325 & 0.000907 & 25.8 & 15047 & 12741 & 1376 \\
     \hline
    \end{tabular} 
\end{center}
 \end{table*}